\documentclass{eas}
\usepackage{astron} 
\usepackage{pdfpages}
\usepackage{graphicx}
\usepackage{psfrag}
\usepackage{pdftricks}
\usepackage{graphicx}
\usepackage{amsmath}
\usepackage{amsfonts}
\usepackage{amssymb}

\begin{document}

\title{MIMAC potential discovery and exclusion of neutralinos in the MSSM and NMSSM}
\runningtitle{MIMAC and neutralinos}

\author{Daniel Albornoz V\'asquez}

\address{LAPTH, U. de Savoie, CNRS, BP 110, 74941 Annecy-Le-Vieux, France}

\begin{abstract}
The MIMAC project aims to provide a nominal fluorine detector for directional detection of galactic dark matter recoil events. Its expected behavior reaches an important part of the predicted spin dependent elastic scattering interactions of the supersymmetric neutralino with protons. Hence, the parameter space in the MSSM and the NMSSM models with neutralino dark matter could be probed by such experimental efforts. In particular, a good sensitivity to spin dependent interactions tackles parameter space regions to which the predictions on spin independent interactions and indirect signatures are far below current and projected experiments.
\end{abstract}

\maketitle

\section{Introduction}
Recent and rapid experimental developments have been constraining particle dark matter (DM) candidates. In particular, the direct detection community has showed its capacity to scan DM-nucleon interactions for candidates of masses at the electroweak scale with spin independent interactions down to $\sim 10^{-44}$ cm$^2$ and spin dependent cross sections down to $\sim 10^{-37}$ cm$^2$ (see~\cite{Censier:2011wd} for a review on the subject).
\\
The directional technique relies on the fact that the solar system is in motion with respect to the galactic reference frame with a velocity pointing toward the Cygnus constellation. The interaction between the DM of the galaxy and a detector on Earth would happen in a preferred direction: recoil events could record this asymmetry. In~\cite{Billard:2009mf,Billard:2010gp} the projected sensitivity to spin dependent proton-DM interactions of a forthcoming fluorine detector is estimated. The characteristics of the simulated detector are set to be those of the MIMAC project, but are also representative of the whole generation of detectors currently in development. Data was simulated and analyzed for a detector made of 10 kg of $\rm CF_4$, operated at 50 mbar, assuming that recoil tracks can be solved, including the head-tail determination of the event, with a threshold at 5 keV, and an exposure of 30 kg yr.
\\
Supersymmetric models provide neutralinos as a DM candidate. The Minimal Supersymmetric Standard Model (MSSM) and Next-to-MSSM (NMSSM)~\cite{Ellwanger:2009dp} are well established realizations of Supersymmetry in which the lightest supersymmetric particle (LSP) is stable. In the MSSM and NMSSM with non-universal gaugino masses one could find neutralinos as light as $\sim 13\; \rm{GeV}$ (see~\cite{Vasquez:2011yq} and references therein) and $\sim 1\; \rm{GeV}$ (see~\cite{Vasquez:2010ru,Draper:2010ew} and references therein) respectively, and still overcome all particle physics constraints. That range being not far from the lower limit of the direct and directional detection sensitivity, it is crucial to make a thorough exploration of the different possibilities, and in particular, to establish the range of elastic scattering and annihilation interactions to be expected.

\section{Directional detection sensitivity to supersymmetric configurations}

\subsection{Exploration of supersymmetric configurations and constraints}
Supersymmetry offers a large number of possibilities in a multidimensional parameter space with new interactions and particles. The phenomenology related to a neutralino DM candidate is mostly determined by its mass and composition, as well as by the Higgs sector. Hence among the relevant free parameters are gaugino masses $\rm M_1$, $\rm M_2$ and $\rm M_3$, the ratio of the vacuum expectation values of the two Higgs doublets $\rm tan\beta$, and the $\rm \mu$ mass term. The MSSM and the NMSSM differ in the Higgs sector, where the NMSSM has an extra singlet scalar field. This gives a natural explanation to the energy scale of the MSSM $\rm \mu$ mass term, which becomes effective in the NMSSM. To complete the Higgs sector parametrization, in the MSSM we have the pseudoscalar mass $\rm M_A$ which is replaced in the NMSSM by the scalar couplings $\rm \lambda$ and $\rm \kappa$, as well as the corresponding trilinear couplings $\rm A_\lambda$ and $\rm A_\kappa$. The Higgs particle spectrum is expanded from the MSSM -with h, H, $\rm H^\pm$ and A- to the NMSSM -with $\rm H_1$, $\rm H_2$, $\rm H_3$, $\rm H^\pm$, $\rm A_1$ and $\rm A_2$. In particular, the stringent constraints on the lightest MSSM scalar Higgs h do not necessarily apply to the lightest NMSSM scalar Higgs $\rm H_1$, since it could be strongly dominated by a singlet component. This, in turn, is at the origin of a broader set of possibilities for physics below the 100 GeV scale in the NMSSM. The set of free parameters is completed by the soft sfermion masses and the trilinear coupling of the top sector A$_t$ -the other trilinear couplings being set to zero. 
\\
Scanning the multidimensional supersymmetric parameter spaces is not an easy task. To this end a Markov Chain Monte-Carlo code was developed to scan the multidimensional supersymmetric parameter spaces~\cite{Vasquez:2010ru}. This code, built on micrOMEGAs 2.4~\cite{Belanger:2005kh}, evaluates each supersymmetric parametrization using a likelihood function to fit particle physics limits, including limits on masses -such as the chargino or the Higgs bosons-, electroweak observables -such as $\rm (g-2)_\mu$ or the Z invisible width-, and B-physics -such as the $\rm b\rightarrow s\gamma$, $\rm B_s \rightarrow \mu^+\mu^-$ and $B\rightarrow \tau \nu_\tau$ branching ratios. Neutralinos are also required to represent at least 10\% of the cosmological DM as measured by WMAP~\cite{Komatsu:2010fb}, hence a likelihood function was established for the relic density of the neutralinos after thermal freeze-out. An iteration process based on a Metropolis-Hastings algorithm generates a random walk in the multidimensional parameter space.
\\
Once the parameter space is explored and configurations allowed by particle physics experiments are found, we may compare the direct and indirect detection yield. Therefore, we compute the spin independent cross sections times neutralino-to-DM fraction $\xi$, to which we apply the exclusion limits by XENON100~\cite{Aprile:2011hi}. We also apply Fermi-LAT constraints on $\gamma$-ray fluxes from the Draco dwarf spheroidal galaxy (dSph)~\cite{Abdo:2010ex}. The $\gamma$-ray flux stemming from neutralino annihilations in the DM dominated Draco galaxy was estimated by computing the $\gamma$-ray production cross section and spectrum from neutralino annihilations, and multiplied by the line-of-sight integral, provided by Fermi-LAT for a Navarro-Frenk-White halo profile~\cite{Navarro:1995iw}. For more details, see~\cite{Vasquez:2011js}.

\subsection{The impact of directional detection on neutralinos}
\begin{figure}[ht]
\centering		
\includegraphics[scale=0.2,natwidth=6cm,natheight=6cm]{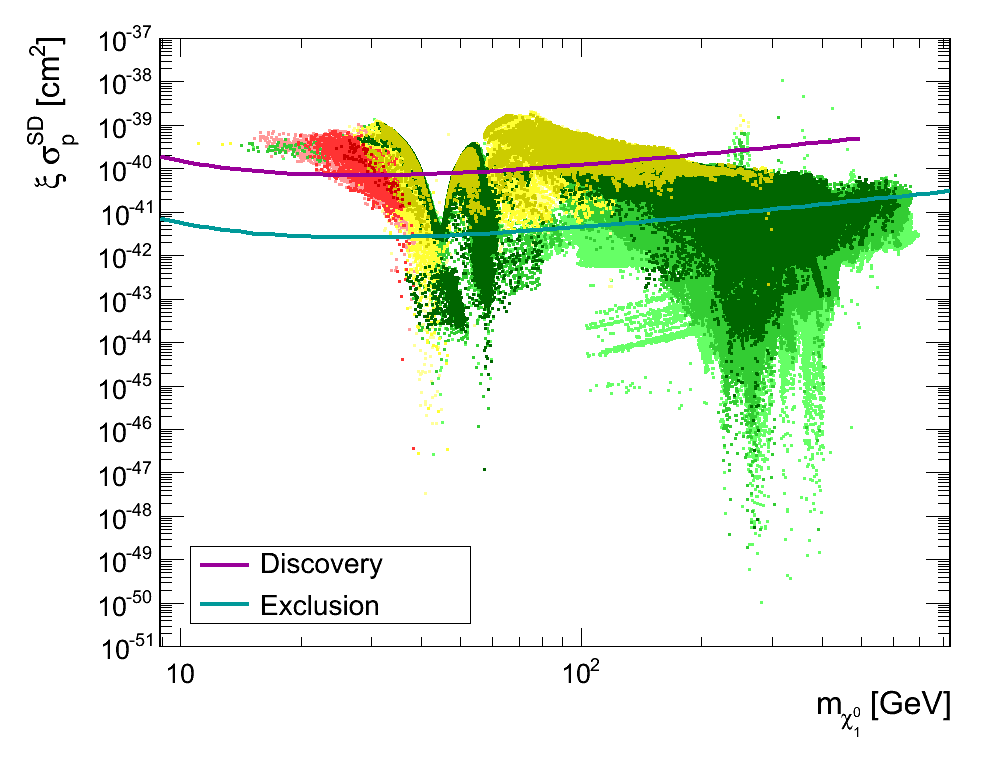}
\includegraphics[scale=0.2,natwidth=6cm,natheight=6cm]{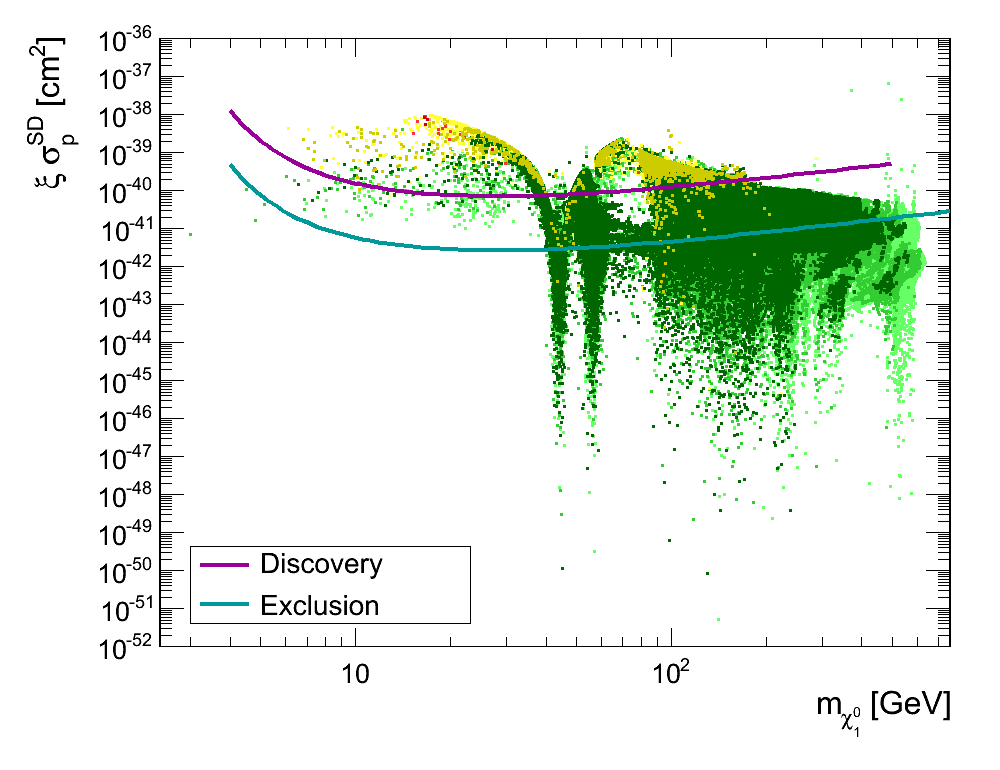}
\caption{Spin dependent proton-neutralino interactions as a function of neutralino mass in the MSSM (top panel) and the NMSSM (bottom panel). The projected sensitivity of a typical directional detector are also shown. In green: points safe with respect to XENON100 limits on spin independent interactions and Fermi-LAT limits on $\gamma$-rays from the Draco dwarf spheroidal galaxy. In yellow: points in conflict with either one or the other. In red: points failing to overcome both exclusion limits.}
\label{fig:DD_SDp_Mchi}
\end{figure}

Directional detectors based on fluorine are sensitive mostly to the spin dependent interactions of DM with protons. The directional detection prospects for a MIMAC-like detector show that if no background events are recorded, the detector would be able to exclude cross section down to $\simeq 4\times 10^{-42}$ cm$^2$ for a mass of 10 GeV. Furthermore, determining the directionality increases the number of observables. Using a likelihood analysis, it is possible to determine the WIMP mass and interaction cross section. The needed statistics to solve these characteristics can be translated into a sensitivity curve, which is, of course, above the exclusion limit. In light of these sensitivity curves, three regions are defined in the $\xi \sigma^{SD}_p$ vs. $m_{\chi_1^0}$ plane ($\xi$ being the neutralino to DM fraction at Earth's position): above the discovery limit, between discovery and exclusion limits, and below the exclusion limits. In the first region, neutralinos are expected to be detected, their mass and cross section could be measured. Of course, if no event is measured, then the corresponding configuration would be ruled out. In the second region, neutralinos would produce some signal but not enough to be solved. If they do not, then scenarios lying in this region would be excluded. Finally the third region is populated by neutralinos that would not yield any effect in the forthcoming directional detection experiments.
\\
We present the results for the MSSM and NMSSM in Fig.~\ref{fig:DD_SDp_Mchi}. It is important to notice that many configurations lie above the projected exclusion limit of MIMAC in both the MSSM and the NMSSM. This already encourages the efforts for building fluorine directional detectors. We show in green points allowed by both XENON100 and Fermi-LAT, while points failing one or the other are tagged in yellow. Those excluded by both are tagged in red.
\\
The predictions for neutralinos in these two models only differ significantly for neutralino masses below $\sim$ 30 GeV, This is a direct consequence of the very different configurations in the Higgs sector. Indeed, as it was found in~\cite{Vasquez:2010ru} and further explored in~\cite{Vasquez:2011js}, NMSSM neutralinos below 30 GeV achieve the relic density by resonantly annihilating via light scalar or pseudoscalar Higgs bosons, absent in the MSSM. In the latter, the only possibilities are to push as much as possible the masses of the lightest scalar Higgs in order to allow lighter pseudoscalar masses, and have efficient enough annihilations. This scenario is actually heavily constrained and should not be regarded as plausible, as shown in~\cite{Vasquez:2011yq}. The other possibility is exchanging very light staus, which provides neutralinos down to 12.6 GeV. Since the MSSM is contained in the NMSSM, this configuration is also realizable in the NMSSM. Hence, a detection below 30 GeV implies different predictions for both models: in the MSSM we would expect to observe light sleptons ($\lesssim 100$ GeV), while the NMSSM could predict a very light Higgs boson.
\\
In both models, there is a prominent concentration of points with $m_{\chi_1^0} \sim$45 GeV, corresponding to neutralinos in resonant annihilations through a Z boson, an efficient mechanism to attain the correct relic density. The points that can be detected around the Z resonance are those not falling exactly in it, but rather those of masses $\lesssim 40$ GeV and $\gtrsim 50$ GeV. This is easily understood in terms of the elastic scattering cross section: the spin dependent interactions occur mainly via the exchange of a Z boson. While the Z resonance represents a good way to obtain a plausible relic density, a too fine-tuned relation between neutralino and Z masses leads to too small abundances unless the coupling to the Z is small. Thus, for points with $m_{\chi_1^0}$ sitting too close to $\rm M_Z/2$, the Z$\chi_1^0\chi_1^0$ coupling is small, hence spin dependent interactions are consequently reduced.
\\
For the larger masses, for LSP with larger higgsino components, the Z$\chi_1^0\chi_1^0$ coupling is usually large, since it is proportional to $N_{13}^2-N_{14}^2$ ($N_{13}^2$ and $N_{14}^2$ being the higgsino-d and higgsino-u fractions of the neutralino). However, above a few hundred GeV the mass split between the lightest neutralino and squarks narrows. Hence both Z and $\tilde{q}_{u,\,d}$ contribute to the interactions. It turns out that these two contributions are destructive. Therefore, for a generally dominating Z exchange with rather large couplings, those configurations having a large enough squark exchange can lower the spin dependent proton-neutralino cross section by a few orders of magnitude. This is why not all the higgsino points have large interactions. Thus, only a fraction of them falls in the discovery region. The general trend to have smaller cross sections towards larger neutralino masses is a consequence of the kinematic behavior of the cross section: when $m_{\chi_1^0}\gg m_{p}$, then the neutralino-proton cross section is proportional to $m_{\chi_1^0}^{-2}$. When the maximum Z$\chi_1^0\chi_1^0$ coupling is achieved -i.e., $N_{13}^2-N_{14}^2\simeq 0.5$-, the upper limit for the interaction cross section as a function of the neutralino mass is the -2 power law observed in both panels in Fig.~\ref{fig:DD_SDp_Mchi}.
\\
Hence, in a very general way, any detection of a neutralino implies a large higgsino fraction, for a neutralino mass below $\sim$ 150 GeV. This implies that the detection of a neutralino fixes the $\mu$ mass term to be lighter than 200 GeV. In turn, this predicts a chargino of mass below 200 GeV, which should be observed at LHC. The other possibility would be small squark masses, which, however, seem to be not an option any more after the first round of results at ATLAS and CMS.
\\
It is also important to notice that many points escape both the discovery region and the exclusion region. This should not be taken as a drawback, but encourage experimental efforts to develop as good techniques as possible. Consequently, it is also important to keep an eye on other detection techniques, in order to tackle the most difficult configurations.

\subsection{Complementarity of direct, indirect and directional detection techniques}
\begin{figure}[ht]
\centering		
\includegraphics[scale=0.2,natwidth=6cm,natheight=6cm]{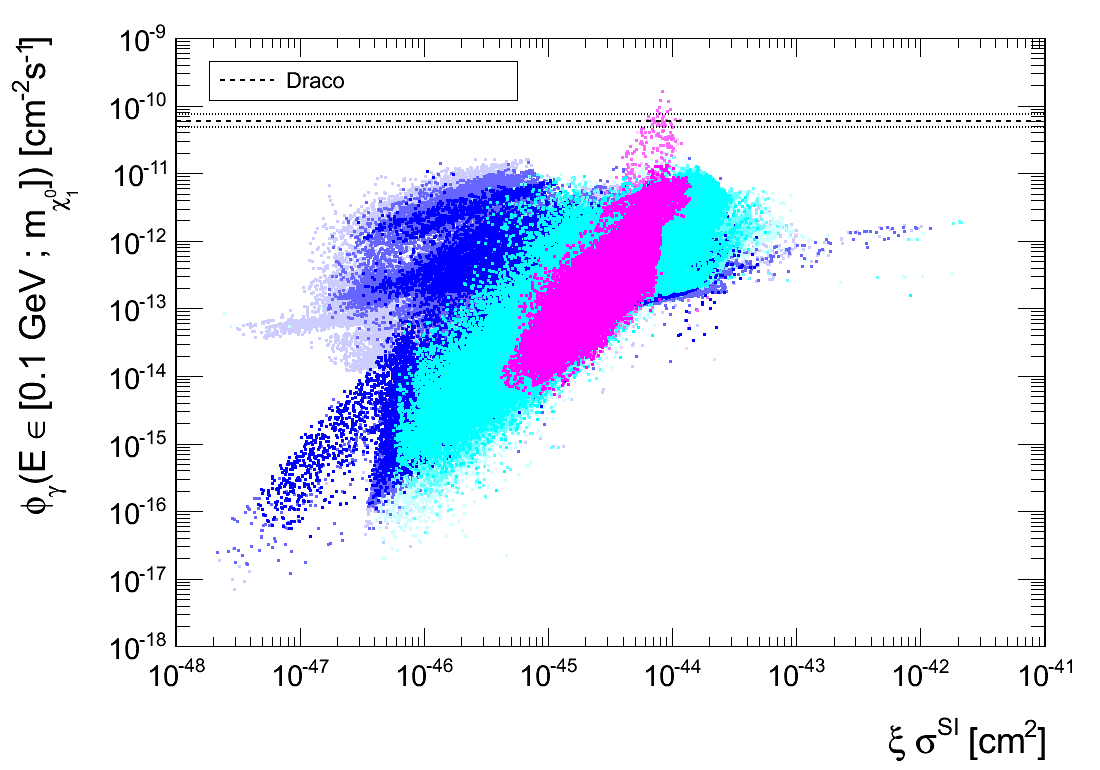}
\includegraphics[scale=0.2,natwidth=6cm,natheight=6cm]{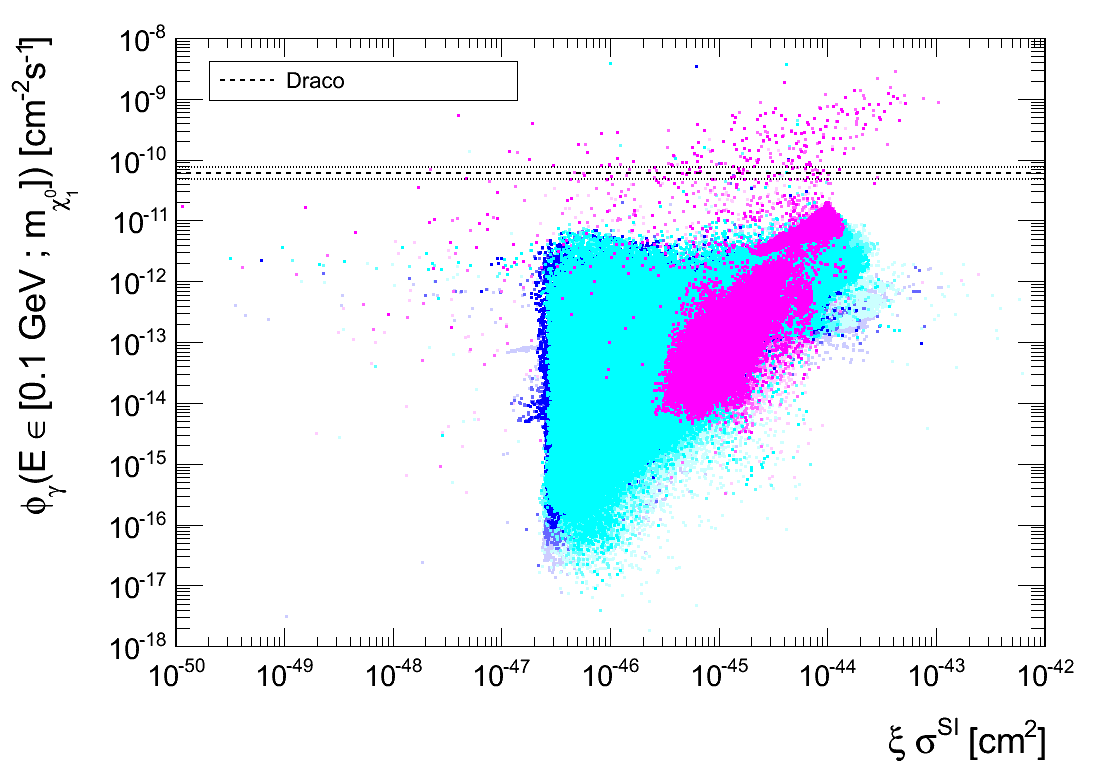}
\caption{Predicted $\gamma$-ray fluxes from the Draco dwarf galaxy as a function of $\xi \sigma^{SI}$ in the MSSM (top panel) and the NMSSM (bottom panel). Points excluded by XENON100 are not drawn. The Fermi-LAT limits for the flux are shown. In magenta: points falling in the MIMAC discovery region. In cyan: points falling in the MIMAC exclusion region. In blue: points beyond the MIMAC sensitivity.}
\label{fig:DD_Draco_SI}
\end{figure}

In Figs.~\ref{fig:DD_Draco_SI} we display the points in the $\gamma$-ray flux vs. $\xi \sigma^{SI}$ plane, using the magenta, cyan and blue tagging for those configurations that could be discovered, excluded or would not yield any effect in a nominal directional detector.
\\
The XENON100 exclusion limits rule out part of the parameter space, which is not shown in these figures. Hence, a fluorine based directional detector would scan configurations that are safe with respect to XENON100 limits on spin independent interactions. Furthermore, in the MSSM, the magenta point with the smallest spin dependent interaction reads $\xi \sigma^{SI} \simeq 4\times10^{-46}$ cm$^2$, which is unlikely to be within the sensitivity of the projected XENON1ton or other spin independent-oriented projected detectors. In the NMSSM this is even more drastic: there is a point in magenta with $\xi \sigma^{SI} \simeq 10^{-50}$ cm$^2$!
\\
Regarding the indirect $\gamma$-ray flux from the Draco dSph, the conclusion is similar: we find discoverable configurations which lie up to four orders of magnitude below the Fermi-LAT limits in the MSSM, and even more in the NMSSM.
\\
It is important for the prospect of directional experiments that we find large concentrations of points which are not excluded by any experiment yet, which are far away from detectability by other techniques such as indirect detection and direct detection, and which could be discovered or excluded by such projected detectors.

\section{Conclusions}
The projected sensitivity for directional detectors such as MIMAC would allow to probe a large portion of parameter space of neutralino DM supersymmetric configurations, especially towards the lightest LSP regions, below 30 GeV. A detection could happen below 150 GeV, and would imply a significant higgsino fraction in the neutralino composition, which in turn predicts a chargino lighter than 200 GeV. For discoveries of even lighter neutralinos, the predictions of the MSSM and the NMSSM are quite different: the former points towards light sleptons while the latter implies light scalar or pseudoscalar Higgs bosons.
\\
The interplay between the projected sensitivity of fluorine directional detectors, the spin independent interactions and the $\gamma$-ray fluxes expected for neutralinos in the MSSM and the NMSSM is a crucial feature for the future explorations of neutralino DM. If the LHC tells us something about Supersymmetry, then we may have indications for which technique is the most adapted to discover or exclude the existence of a neutralino in galactic systems. Conversely, signals in direct, directional or indirect detection could help the LHC to confirm or rule out the MSSM and/or the NMSSM.

\bibliography{Cygnus_Directional_Susy}

\begin{thebibliography}{}

\bibitem[\protect\astroncite{Abdo et~al.}{2010}]{Abdo:2010ex}
Abdo, A., Ackermann, M., Ajello, M., Atwood, W., Baldini, L., et~al.: 2010,
\newblock {\em Astrophys.J.} {\bf 712}, 147

\bibitem[\protect\astroncite{Albornoz~Vasquez et~al.}{2011a}]{Vasquez:2011js}
Albornoz~Vasquez, D., Belanger, G., and Boehm, C.: 2011a,
\newblock arXiv:1107.1614

\bibitem[\protect\astroncite{Albornoz~Vasquez et~al.}{2011b}]{Vasquez:2011yq}
Albornoz~Vasquez, D., Belanger, G., and Boehm, C.: 2011b,
\newblock arXiv:1108.1338

\bibitem[\protect\astroncite{Albornoz~Vasquez et~al.}{2010}]{Vasquez:2010ru}
Albornoz~Vasquez, D., Belanger, G., Boehm, C., Pukhov, A., and Silk, J.: 2010,
\newblock {\em Phys.Rev.} {\bf D82}, 115027

\bibitem[\protect\astroncite{Aprile et~al.}{2011}]{Aprile:2011hi}
Aprile, E. et~al.: 2011,
\newblock {\em Phys.Rev.Lett.}

\bibitem[\protect\astroncite{Belanger et~al.}{2005}]{Belanger:2005kh}
Belanger, G., Boudjema, F., Hugonie, C., Pukhov, A., and Semenov, A.: 2005,
\newblock {\em JCAP} {\bf 0509}, 001

\bibitem[\protect\astroncite{Billard et~al.}{2010a}]{Billard:2009mf}
Billard, J., Mayet, F., Macias-Perez, J., and Santos, D.: 2010a,
\newblock {\em Phys.Lett.} {\bf B691}, 156

\bibitem[\protect\astroncite{Billard et~al.}{2010b}]{Billard:2010gp}
Billard, J., Mayet, F., and Santos, D.: 2010b,
\newblock {\em Phys.Rev.} {\bf D82}, 055011

\bibitem[\protect\astroncite{Censier}{2011}]{Censier:2011wd}
Censier, B.: 2011,
\newblock {\em ibid.},
\newblock arXiv:1110.0191

\bibitem[\protect\astroncite{Draper et~al.}{2011}]{Draper:2010ew}
Draper, P., Liu, T., Wagner, C.~E., Wang, L.-T., and Zhang, H.: 2011,
\newblock {\em Phys.Rev.Lett.} {\bf 106}, 121805

\bibitem[\protect\astroncite{Ellwanger et~al.}{2010}]{Ellwanger:2009dp}
Ellwanger, U., Hugonie, C., and Teixeira, A.~M.: 2010,
\newblock {\em Phys.Rept.} {\bf 496}, 1

\bibitem[\protect\astroncite{Komatsu et~al.}{2011}]{Komatsu:2010fb}
Komatsu, E. et~al.: 2011,
\newblock {\em Astrophys.J.Suppl.} {\bf 192}, 18

\bibitem[\protect\astroncite{Navarro et~al.}{1996}]{Navarro:1995iw}
Navarro, J.~F., Frenk, C.~S., and White, S.~D.: 1996,
\newblock {\em Astrophys.J.} {\bf 462}, 563

\end{thebibliography}
\bibliographystyle{astron}

\end{document}